%% file: main.tex
\documentclass[11pt]{article}

\usepackage[utf8]{inputenc}
\usepackage[T1]{fontenc}
\usepackage[margin=1in]{geometry}
\usepackage{amsmath,amssymb}
\usepackage{graphicx}
\usepackage{booktabs}
\usepackage{tabularx}
\usepackage{float}
\usepackage{enumitem}

\usepackage{cite}

\usepackage{hyperref}
\usepackage{xurl}

\hypersetup{
    colorlinks=true,
    linkcolor=black,
    citecolor=black,
    urlcolor=blue,
    pdftitle={Trusted Credentials, Untrusted Behavior: Benchmarking LLM-Agent Security in High-Performance Computing},
    pdfauthor={Jie Li}
}

\title{Trusted Credentials, Untrusted Behavior: \\
  Benchmarking LLM-Agent Security in High-Performance Computing
  \thanks{Position/vision paper.}}

\author{
  Jie Li \\
  \small Texas Tech University \\
  \small \href{mailto:jie.li@ttu.edu}{jie.li@ttu.edu}
}

\date{\today}

\begin{document}

\maketitle

\input{sections/00_abstract}

\input{sections/01_introduction}
\input{sections/02_background}
\input{sections/03_gap}
\input{sections/04_threat_model}
\input{sections/05_taxonomy}
\input{sections/06_defense_gap}
\input{sections/07_research_agenda}
\input{sections/08_conclusion}

\bibliographystyle{unsrt}
\bibliography{references}

\end{document}

%% file: sections/00_abstract.tex
\begin{abstract}
\noindent
Large language model (LLM) agents are starting to take on routine work in high-performance computing (HPC), including monitoring Slurm jobs, diagnosing failed builds, inspecting simulation output, and coordinating scientific workflows. To do this work, an agent commonly acts under its user's credentials and inherits the user's access to files and the scheduler. This arrangement creates a failure mode that ordinary account-level controls do not capture. Adversarial instructions in a log, tool description, shared file, or peer-agent message may redirect the agent beyond the task the user assigned, even though every resulting command is authenticated and permitted for that account. We refer to this as the \emph{hijacked authorized agent} problem. Existing agent-security studies explain relevant mechanisms, such as indirect prompt injection and tool misuse, but generally evaluate them in web, enterprise, or personal-assistant settings. HPC security, by contrast, has mature controls for identity and isolation but does not ordinarily represent the intent of a particular task. This paper defines the threat model in the HPC setting, identifies attack surfaces created by schedulers, shared storage, multi-project accounts, and scientific workflows, and examines where current controls fall short. It concludes with a research agenda and a plan for an empirical benchmark, TaskBound.
\end{abstract}

%% file: sections/01_introduction.tex
\section{Introduction}
\label{sec:intro}

Much of the effort involved in using an HPC system happens around the batch job itself. Users inspect logs, repair scripts, install dependencies, sort through output, and decide what to run next. These are natural tasks for an LLM agent with access to files, command-line tools, and scheduler commands. Several recent systems connect agents to HPC resources and scientific workflows \cite{Ma2025HPCAgent,Rosendo2025ORNL,Dawson2026LARA}; another uses multiple agents to build and repair HPC software \cite{Mondesire2025PEARC}. Adoption is also occurring outside research prototypes. Aalto Scientific Computing, for example, has published guidance for users running agentic coding tools on its Triton cluster \cite{AaltoSciComp2026}.

Giving an agent enough access to be useful also gives untrusted material a new route to that access. A debugging agent must read the output of a failed job. A build agent must interpret compiler messages and package metadata. A workflow agent may exchange files or messages with other agents. If any of this material contains adversarial instructions, the agent may act on them using the authority it received from the user. The resulting commands can be valid from the operating system's point of view and still be outside the user's intended task.

Agent-security research provides a good account of the underlying mechanisms. Indirect prompt injection can manipulate applications that consume untrusted content \cite{Greshake2023PromptInjection}; tool-using agents can take harmful actions \cite{Ruan2024ToolEmu}; and benchmarks can measure both task completion and resistance to attack \cite{Debenedetti2024AgentDojo}. Tool descriptions and inter-agent protocols add further avenues for manipulation \cite{Shi2025ToolHijacker,Ferrag2025ProtocolExploits,Wang2026PILandscape}. Most evaluations, however, use settings such as email, banking, travel, or generic tool sandboxes. They do not model scheduler allocations, parallel filesystems, or accounts that legitimately span several scientific projects.

HPC security starts from a different concern: deciding which principals may use which resources. Zone-based architectures separate access, management, compute, and storage functions \cite{NIST2024SP800223}. Deployed systems isolate users' processes, files, network traffic, and accelerators \cite{Prout2024UserSeparation}, while federated identity and zero-trust designs strengthen authentication and access control across facilities \cite{Alam2024ZeroTrust}. These controls remain necessary. Yet an access check cannot determine whether a permitted command reflects the user's request or an instruction copied from a poisoned log into the agent's context.

Our concern lies between these two lines of work. The identity and credentials may be valid, and each action may pass the site's authorization checks, while the sequence of actions no longer serves the task the user assigned. We call an agent in this state a \emph{hijacked authorized agent}.

\subsection{Position}
\label{sec:thesis}

Our position rests on three observations:

\begin{enumerate}[leftmargin=*]
  \item HPC changes the practical meaning of a successful prompt-injection attack. Scheduler authority, durable shared storage, multi-project access, and scientific provenance introduce consequences that are absent from many current agent benchmarks.
  \item Neither neighboring literature measures the complete problem. Agent-security benchmarks omit important HPC resources and policies, while HPC controls do not test whether adversarial content can redirect an authenticated agent.
  \item The question is already relevant to deployed systems. Agents are being used in HPC workflows \cite{Ma2025HPCAgent,Rosendo2025ORNL,Mondesire2025PEARC}, and at least one computing center has begun documenting the associated risks \cite{AaltoSciComp2026}. Evaluation should develop alongside adoption rather than after local practices have hardened.
\end{enumerate}

\subsection{Contributions}
\label{sec:contributions}

\begin{enumerate}[leftmargin=*]
  \item We define the \textbf{hijacked authorized agent} threat model and distinguish it from privilege escalation and generic indirect prompt injection (Section~\ref{sec:threat-model}).
  \item We organize the relevant HPC attack surfaces into five groups: shared-filesystem poisoning, scheduler and log channels, tool/module/MCP poisoning, cross-project leakage, and multi-agent exfiltration (Section~\ref{sec:taxonomy}).
  \item We compare these surfaces with existing HPC and agent-monitoring controls (Section~\ref{sec:defense-gap}).
  \item We identify open research problems and describe TaskBound, a companion benchmark now under development (Section~\ref{sec:research-agenda}).
\end{enumerate}

\subsection{Scope}
\label{sec:scope}

This is a threat characterization and research agenda, not an empirical evaluation. We do not report attack success rates, assess a particular model or framework, or present a complete defense. Training-time poisoning, weight extraction, GPU side channels, kernel or hypervisor compromise, and general content-safety jailbreaks are also outside our scope. The narrower question here is how content from the HPC environment can redirect an agent that already holds valid credentials.

%% file: sections/02_background.tex
\section{Background}
\label{sec:background}

\subsection{LLM agents are entering HPC}
\label{sec:background-hpc}

Several projects now place LLM agents in direct contact with HPC infrastructure. Ma et al.\ connect LangChain and LangGraph agents to HPC resources through Parsl and demonstrate molecular-dynamics workflows on ALCF's Polaris system \cite{Ma2025HPCAgent}. Rosendo et al.\ combine an LLM interface, multi-agent decision making, facility APIs, and provenance services for experiments spanning two ORNL user facilities \cite{Rosendo2025ORNL}. Mondesire et al.\ use a multi-agent system to build and repair more than 200 HPC software packages \cite{Mondesire2025PEARC}. In the last example, compiler errors and build logs are not incidental output; they are inputs to the next decision made by the agent.

Dawson et al.\ identify correctness, reproducibility, and safe resource use as open problems for agentic supercomputing workflows \cite{Dawson2026LARA}. Similar concerns have reached site documentation. Aalto Scientific Computing notes the use of Claude Code and OpenAI Codex for coding and Slurm job management on Triton, and warns users about prompt injection, malicious skills, external code, and confidentiality \cite{AaltoSciComp2026}. The systems literature nevertheless remains focused mainly on capability, orchestration, and provenance. It offers little evidence about behavior when routine inputs such as logs or tool output are malicious.

\subsection{LLM-agent security: known attack mechanisms}
\label{sec:background-agent-security}

The underlying attack mechanisms have been studied mostly outside HPC. Greshake et al.\ showed that instructions embedded in retrieved content, rather than supplied by the user, can change an LLM-integrated application's behavior and lead to data disclosure or persistent compromise \cite{Greshake2023PromptInjection}. ToolEmu evaluates harmful actions by tool-using agents in an LM-emulated sandbox \cite{Ruan2024ToolEmu}. AgentDojo provides a reproducible benchmark with 97 tasks and 629 security test cases across email, banking, travel, and workspace applications \cite{Debenedetti2024AgentDojo}. Its joint measurement of task utility and attack success is particularly useful for our proposed benchmark (Section~\ref{sec:benchmark}).

Recent work broadens the picture. Ferrag et al.\ survey attacks that include inter-agent protocol exploits \cite{Ferrag2025ProtocolExploits}, and Wang et al.\ systematize prompt-injection threats and their defenses \cite{Wang2026PILandscape}. Shi et al.\ show that a malicious tool description can steer an agent toward an attacker-controlled tool without requiring white-box access to the model \cite{Shi2025ToolHijacker}. This result is relevant to environment modules, shared utilities, and MCP tool servers (Section~\ref{sec:tool-poisoning}). Still, an email inbox or travel itinerary does not exercise scheduler authority, project allocations, or filesystem state shared across jobs and users. Those differences affect both the available attacks and the definition of a policy violation.

\subsection{HPC security: established controls and their assumptions}
\label{sec:background-hpcsec}

HPC security has developed strong controls around identities and resources. NIST SP 800-223 organizes HPC infrastructure into access, management, compute, and storage zones and analyzes threats to each \cite{NIST2024SP800223}. Prout et al.\ describe user separation at the MIT Lincoln Laboratory Supercomputing Center across processes, files, network traffic, and accelerators \cite{Prout2024UserSeparation}. Alam et al.\ present a federated single-sign-on and zero-trust design for AI and HPC research infrastructure, including multi-factor authentication and time-limited role-based access \cite{Alam2024ZeroTrust}.

These mechanisms determine who may access a resource and under what conditions. They generally treat a successful authorization decision as sufficient evidence that the action is allowed. For an LLM agent, that inference is incomplete. A POSIX check or scheduler record cannot tell whether the user requested an action or whether the agent derived it from adversarial text in a log. The site sees the same account in either case. The missing information is the scope and origin of the task that led to the action.

%% file: sections/03_gap.tex
\section{What Existing Work Leaves Open}
\label{sec:gap}

The three bodies of work reviewed above stop at different boundaries. Agentic-HPC systems demonstrate useful access to schedulers, files, and scientific instruments, but seldom evaluate hostile inputs encountered during a workflow. Agent-security benchmarks do evaluate hostile inputs, though without HPC resources or policies. HPC security protects those resources, but its enforcement point is usually the user or account rather than the narrower task delegated to an agent.

What remains poorly characterized is therefore a common and plausible case: an agent holds valid HPC credentials, reads content in the course of authorized work, and is redirected into a different but still account-permitted action. We have not found prior work that treats this intersection as its primary object of study. The rest of the paper gives the problem a threat model, identifies its HPC-specific attack surfaces, and asks what an evaluation would need to measure.

%% file: sections/04_threat_model.tex
\section{Threat Model: The Hijacked Authorized Agent}
\label{sec:threat-model}

\subsection{Definition}
\label{sec:threat-def}

We call an agent \textbf{hijacked} when content encountered during ordinary operation causes it to act outside its assigned task, even though the user's account and the agent's tool configuration permit each individual action. The content might appear in a file, a job log, tool output, or a message from another agent. No conventional privilege escalation is required: the attacker need not obtain root access, bypass POSIX permissions, or steal credentials.

Consider an agent asked to diagnose a failed job in project A. The user may also belong to project B and may be allowed to submit jobs to several partitions. If a poisoned log leads the agent to read project B or launch an unrelated job, the operating system and scheduler may record nothing more than permitted activity by that user. The violated boundary is the authority implied by the debugging task, not the standing authority of the account. Conventional HPC controls do not usually encode that task boundary in a form they can enforce or audit.

\subsection{Assumptions}
\label{sec:threat-assumptions}

\begin{itemize}[leftmargin=*]
  \item the user intentionally and legitimately launches the agent for a specific task;
  \item the agent has valid, correctly provisioned access to its assigned workspace and tools;
  \item the cluster control plane, scheduler, and any benchmark or monitoring harness are trusted;
  \item the attacker does not have root access and does not modify model weights;
  \item the attacker's only foothold is the ability to influence content the agent will read, execute, or receive during normal operation, such as a file, a log, a tool's output, shared state, or a message from another agent.
\end{itemize}

\subsection{Attacker capability classes}
\label{sec:threat-capabilities}

Table~\ref{tab:capabilities} lists the attacker footholds considered here. Stating the foothold matters because an attack that already assumes unauthorized filesystem or administrator access demonstrates a conventional access-control failure, not the failure mode defined above.

\begin{table}[H]
\centering
\caption{Attacker capability classes considered by the hijacked authorized agent threat model.}
\label{tab:capabilities}
\begin{tabularx}{\linewidth}{@{} l X X @{}}
\toprule
\textbf{Class} & \textbf{Capability} & \textbf{Representative HPC source} \\
\midrule
C1 & Write or influence a scientific artifact the agent will read & simulation log, metadata field, README, result file \\
C2 & Influence scheduler-adjacent output & job stdout/stderr, wrapper diagnostics, accounting note \\
C3 & Control a tool's description or output & build helper, converter, module description, MCP tool server \\
C4 & Write to shared node or filesystem state & scratch space, cache, shared collaboration path \\
C5 & Control or hijack a peer agent & inter-agent message, staged intermediate artifact \\
\bottomrule
\end{tabularx}
\end{table}

%% file: sections/05_taxonomy.tex
\section{Attack Surfaces in HPC}
\label{sec:taxonomy}

Prompt injection, poisoned tool output, and protocol abuse describe how an agent is manipulated. For HPC, we must also ask where the malicious content enters the workflow and what the redirected agent can affect. The following taxonomy is organized around those HPC surfaces. Its categories can lead to different harms, including data disclosure, allocation abuse, and corruption of scientific results, so they should not be treated as interchangeable instances of prompt injection.

\subsection{Shared parallel-filesystem poisoning}
\label{sec:taxonomy-filesystem}

Parallel filesystems and scratch spaces preserve state across jobs and, depending on site policy, expose shared or collaborative paths to several users. Malicious text placed in a result file, cache, README, or predictable output path may remain until a later agent reads it. The writer and reader need not run at the same time, which makes this a persistent workflow channel rather than a one-session interaction.

\subsection{Scheduler-induced co-location and log injection}
\label{sec:taxonomy-scheduler}

Agents that diagnose failures routinely inspect standard output, standard error, accounting records, and wrapper diagnostics. These inputs cannot simply be discarded as untrusted because they contain the evidence needed to repair a job. If an attacker can influence a related job or a shared output path, the resulting text enters a context in which scheduler commands are likely to be available. A successful redirection could therefore consume an allocation by submitting, resizing, or repeatedly resubmitting jobs.

\subsection{Tool, module, and MCP poisoning}
\label{sec:tool-poisoning}

HPC software stacks include site modules, group-maintained build helpers, post-processing scripts, and, increasingly, MCP tool servers. Trust is often based on a mixture of site policy and local convention. Shi et al.\ show that a crafted tool description, not only executable output, can bias an agent's tool selection \cite{Shi2025ToolHijacker}. In an HPC workflow, a module description or MCP manifest may therefore influence the agent before the selected program runs.

\subsection{Cross-project data leakage}
\label{sec:cross-project}

Researchers often belong to several projects, with access to separate allocations and datasets. An agent working on one project inherits whatever broader access the user exposes to it. A prompt-injection attack can exploit this mismatch by directing the agent to read from another project and disclose the result through an otherwise approved channel. Account-level permissions will not flag the read when the user is a legitimate member of both projects.

\subsection{Coordinated multi-agent exfiltration}
\label{sec:taxonomy-multiagent}

Multi-agent workflows pass messages and intermediate artifacts between planners and workers \cite{Rosendo2025ORNL,Mondesire2025PEARC}. If one agent or artifact is compromised, it can steer a second agent that has different access. The resulting disclosure may be spread across several ordinary-looking actions: one agent writes an artifact, another reads protected data, and a third sends context to an approved model backend. Detecting the sequence requires more than reviewing each agent's log in isolation.

%% file: sections/06_defense_gap.tex
\section{Limits of Existing Controls}
\label{sec:defense-gap}

Table~\ref{tab:defense-gap} compares established HPC and agent-monitoring controls with the information needed to recognize a hijacked agent. These controls still reduce risk; the issue is that they operate at a different boundary. Authentication, permissions, quotas, and isolation describe what an account may do. They do not record what a particular delegated task was meant to do. Accounting logs, ACLs, and network allowlists therefore cannot reconstruct task intent after an incident unless the agent or its runtime records that intent and links it to subsequent actions.

\begin{table}[H]
\centering
\caption{Information captured by existing controls and information needed to recognize a hijacked agent.}
\label{tab:defense-gap}
\begin{tabularx}{\linewidth}{@{} X X X @{}}
\toprule
\textbf{Existing control} & \textbf{What it verifies} & \textbf{What it misses for hijacked agents} \\
\midrule
Authentication / SSO / zero-trust identity \cite{Alam2024ZeroTrust} & The identity and attributes of the user & Whether an authenticated action originated in the user's request or in injected content \\
POSIX permissions, zone-based access architecture \cite{NIST2024SP800223} & Resources the account may access & Whether those resources belong to the current task \\
Scheduler accounting and quotas & Resource use attributed to an account & Whether that use was requested or induced by poisoned content \\
Enforced user/process/network separation \cite{Prout2024UserSeparation} & Isolation between users & Redirection within one user's authorized session \\
Filesystem/network auditing, DLP & Data movement over observed channels & Disclosure through an approved channel, such as a normal LLM API request \\
Per-agent action logging & The actions selected by the agent & Whether an action follows task intent or adversarial context \\
\bottomrule
\end{tabularx}
\end{table}

%% file: sections/07_research_agenda.tex
\section{Research Agenda}
\label{sec:research-agenda}

\subsection{Security requirements for HPC-agent platforms}
\label{sec:requirements}

The comparison in Section~\ref{sec:defense-gap} suggests several capabilities for an HPC-agent platform. Some belong in the agent runtime and others may require support from the scheduler, storage system, or site network. Their feasibility will vary across facilities, but together they define a useful design target:

\begin{itemize}[leftmargin=*]
  \item \textbf{Task-scoped authority.} The runtime should distinguish resources needed for the current task from resources available to the user's account as a whole.
  \item \textbf{Context provenance.} User instructions, file content, tool output, and peer-agent messages should retain source and trust metadata when they enter the agent's context.
  \item \textbf{Egress control.} Agent-to-model traffic needs policy enforcement because an approved, encrypted LLM API request can also carry protected data.
  \item \textbf{Shared-filesystem hygiene.} Sites need ways to limit or label stale and cross-user content in scratch, cache, and collaboration paths used by agents.
  \item \textbf{Tool and module trust.} Software-supply-chain checks should cover tool descriptions and MCP manifests as well as executable code.
  \item \textbf{Scheduler-aware containment.} Limits should prevent a redirected agent from turning a context compromise into excessive spending of the user's allocation.
  \item \textbf{Cross-agent correlation.} Monitoring should connect related actions across agents so that a staged disclosure does not appear as several unrelated events.
\end{itemize}

\subsection{Toward a benchmark: TaskBound}
\label{sec:benchmark}

The claims above need empirical support. We are developing a companion benchmark called \textbf{TaskBound} to translate the threat model into executable tasks and attacks. The design follows AgentDojo's practice of measuring task utility together with attack success \cite{Debenedetti2024AgentDojo}, but replaces web and office applications with HPC-specific resources. Planned scenarios include Slurm actions, project-scoped filesystem policies, and checks for silent changes to scientific parameters, filters, or provenance. Each scenario pairs a user task with an explicit task policy and a deterministic security oracle. This should make it possible to compare agents and defenses on three practical questions: whether the agent completes the assigned work, whether adversarial context redirects it, and whether a defense improves security without making the workflow unusable.

\subsection{Open questions}
\label{sec:open-questions}

The benchmark will address only part of the larger problem. Several design questions also deserve study:

\begin{itemize}[leftmargin=*]
  \item How should users specify task scope when legitimate scientific work is exploratory, and how sensitive are security results to that specification?
  \item Do provenance labels and tool allowlists developed for web or enterprise agents transfer to scheduler and filesystem operations?
  \item Should task policy be enforced by the agent framework, scheduler, filesystem, or a separate monitor, especially at sites that cannot modify all four?
  \item How can a multi-agent workflow preserve useful delegation while preventing one agent from silently extending another agent's authority?
\end{itemize}

%% file: sections/08_conclusion.tex
\section{Conclusion}
\label{sec:conclusion}

LLM agents already interact with HPC software, filesystems, and schedulers \cite{Ma2025HPCAgent,Rosendo2025ORNL,Mondesire2025PEARC,AaltoSciComp2026}. Their usefulness depends on reading the same logs, tools, and shared artifacts that can carry adversarial content. When such content redirects an agent, account-level security may see only valid credentials issuing permitted commands. We use \emph{hijacked authorized agent} for this failure mode and argue that its defining boundary is the delegated task rather than the user account. The taxonomy and control analysis in this paper provide a starting point, but measurement is still missing. TaskBound is intended to supply that evidence by testing agents and defenses against realistic scheduler, storage, and scientific-integrity policies. The immediate question is not whether all HPC agents are unsafe. It is whether deployments can state what an agent was allowed to do and detect when its behavior departs from that task.

\subsection*{AI Disclosure}
AI-assisted tools were used to identify literature, check citations, and support drafting. The authors reviewed and revised the manuscript and take responsibility for its claims and conclusions.